\documentclass[fleqn]{2020SCGE}
\usepackage{hyperref}

\usepackage{multirow}
\usepackage{amssymb}
\usepackage{ulem}

\newcommand{\mpch}{\>h^{-1}{\rm {Mpc}}}

\newcommand{\bmu}{{\bar{\mu}}}
\newcommand{\bmxi}{{\bm \xi}}

\usepackage[T1]{fontenc}
\usepackage{graphicx}	
\usepackage{amsmath}	
\usepackage{amssymb}	
\usepackage{bm}

\UseRawInputEncoding

\begin{document}
\ensubject{subject}

\ArticleType{Article}
\SpecialTopic{SPECIAL TOPIC: }
\Year{2023}
\Month{xxx}
\Vol{xx}
\No{x}
\DOI{xx}
\ArtNo{000000}
\ReceiveDate{xxx xxx, 2023}
\AcceptDate{xxx xxx, 2023}

\newcommand{\zd}[1]{\textcolor{red}{[\textbf{ZD}: #1}]}
\newcommand{\csaulder}[1]{\textbf{\color{magenta}{{#1}}}}

\title{Cosmological distance forecasts for the CSST Galaxy Survey using BAO peaks}{}
\author[1,2]{Feng Shi}{{fshi@xidian.edu.cn}}
\author[1,2]{Jieyi Tian}{}
\author[3]{Zhejie Ding}{}
\author[4,5]{Xiaohu Yang}{}
\author[4,5]{Yizhou Gu}{}
\author[6,7]{Christoph Saulder}{}
\author[1,2]{\\Xiaoping Li}{}
\author[1,2]{Yanming Liu}{}
\author[1,2]{Zitong Wang}{}
\author[8,9]{Hu Zhan}{}
\author[8]{Ming Li}{}
\author[10]{Xiaolei Li}{}
\author[11]{\\Hong Guo}{}
\author[8]{Yan Gong}{}
\author[12]{Yunkun Han}{}
\author[13]{Cheng Li}{}
\author[3,4]{Yipeng Jing}{}
\author[8]{\\Jipeng Sui }{}
\author[14,15]{Run Wen}{}
\author[8]{Gong-Bo Zhao}{}
\author[8]{Hu Zou}{}
\author[3,4,5]{Pengjie Zhang}{}
\author[16,14]{\\Xianzhong Zheng}{}
\author[8]{Xingchen Zhou}{}

\AuthorMark{Shi et al.}

\AuthorCitation{Shi, F. et al.}

\address[1]{School of Aerospace Science and Technology, Xidian University, Xi'an 710126, China}
\address[2]{Shaanxi Key Laboratory of Space Extreme Detection, Xidian University, Xi'an 710126, China}
\address[3]{Department of Astronomy, School of Physics and Astronomy, Shanghai Jiao Tong University, Shanghai 200240, China}
\address[4]{Tsung-Dao Lee Institute  \& School of Physics and Astronomy, 
 Shanghai Jiao Tong University, Shanghai 200240, China}
\address[5]{Key Laboratory for Particle Physics, Astrophysics and Cosmology (MoE), and Shanghai Key Laboratory for Particle Physics and Cosmology, \\
Shanghai Jiao Tong University, Shanghai {\rm 200240}, China}
\address[6]{Max Planck Institute for Extraterrestrial Physics, Garching 85748, Germany}
\address[7]{Universit\"ats-Sternwarte M\"unchen,  Munich 81679, Germany}
\address[8]{National Astronomical Observatories, Chinese Academy of Sciences, Beijing 100101, China}
\address[9]{   The Kavli Institute for Astronomy and Astrophysics, Peking University, Beijing 100871, China}
\address[10]{  College of Physics, Hebei Normal University, Shijiazhuang 050024, China}
\address[11]{  Key Laboratory for Research in Galaxies and Cosmology, Shanghai Astronomical Observatory, Shanghai 200030, China}
\address[12]{  Yunnan Observatories, Chinese Academy of Sciences, Kunming 650216, China}
\address[13]{  Department of Astronomy, Tsinghua University, Beijing 100084, China}
\address[14]{  Purple Mountain Observatory, Chinese Academy of Sciences,, Nanjing, 210023, China}
\address[15]{  School of Astronomy and Space Sciences, University of Science and Technology of China, Hefei 230026, China}
\address[16]{  Tsung-Dao Lee Institute and Key Laboratory for Particle Physics, Astrophysics and Cosmology, Ministry of Education, \\  Shanghai Jiao Tong University, Shanghai 201210, China}



\abstract{The measurement of cosmological distances using baryon acoustic oscillations (BAO) is crucial for studying the universe's expansion. The Chinese Space Station Telescope (CSST) galaxy redshift survey, with its vast volume and sky coverage, provides an opportunity to address key challenges in cosmology. However, redshift uncertainties in galaxy surveys can degrade both angular and radial distance estimates. In this study, we forecast the precision of BAO distance measurements using mock CSST galaxy samples, applying a two-point correlation function (2PCF) wedge approach to mitigate redshift errors. We simulate redshift uncertainties of $\sigma_0 = 0.003$ and $\sigma_0 = 0.006$, representative of expected CSST errors, and examine their effects on the BAO peak and distance scaling factors, $\alpha_\perp$ and $\alpha_\parallel$, across redshift bins within $0.0 < z \leqslant 1.0$. The wedge 2PCF method proves more effective in detecting the BAO peak compared to the monopole 2PCF, particularly for $\sigma_0 = 0.006$. Constraints on the BAO peaks show that $\alpha_\perp$ is well constrained around 1.0, regardless of $\sigma_0$, with precision between 1\% and 3\% across redshift bins. In contrast, $\alpha_\parallel$ measurements are more sensitive to increases in $\sigma_0$. For $\sigma_0 = 0.003$, the results remain close to the fiducial value, with uncertainties ranging between 4\% and 9\%; for $\sigma_0 = 0.006$, significant deviations from the fiducial value are observed. We also study the ability to measure parameters $(\Omega_m, H_0r_\mathrm{d})$ using distance measurements, proving robust constraints as a cosmological probe under CSST-like redshift uncertainties. These findings demonstrate that the CSST survey enables few-percent precision measurements of $D_A$ using the wedge 2PCF method, highlighting its potential to place tight constraints on the universe's expansion history and contribute to high-precision cosmological studies.}



\maketitle


\begin{multicols}{2}
\section{Introduction}\label{sec:intro}

Understanding the accelerated expansion of the universe remains a central challenge in cosmology. One of the most powerful observational tools in addressing this challenge is Baryon Acoustic Oscillations (BAO), a relic imprint from the early universe (e.g., see the review by \cite{2013PhR...530...87W}). BAO represents oscillatory patterns in the distribution of galaxies, driven by acoustic waves in the photon-baryon plasma before the epoch of recombination. As the universe expanded, these oscillations became imprinted in the matter distribution, leaving a characteristic scale in both the galaxy correlation function and power spectrum. By measuring this BAO scale, precise cosmological distances can be extracted, establishing BAO as a robust standard ruler for probing the universe's expansion history. Along with other probes, such as supernovae \cite{1998AJ....116.1009R, 1999ApJ...517..565P} and cosmic microwave background (CMB) observations (e.g. \cite{2013ApJS..208...19H, 2020A&A...641A...6P}), BAO measurements have imposed stringent constraints on cosmological models, particularly regarding dark energy.

Galaxy redshift surveys have been essential for detecting the BAO signal and refining constraints on cosmological parameters. The BAO feature was first observed in galaxy clustering by the SDSS \cite{2005ApJ...633..560E} and 2dFGRS \cite{2005MNRAS.362..505C} surveys. Subsequent large-scale surveys, such as the 6dF Redshift Survey  \cite{2011MNRAS.416.3017B}, WiggleZ Dark Energy Survey \cite{2012MNRAS.425..405B},  as well as the Baryon Oscillation Spectroscopic Survey (BOSS) \cite{2017MNRAS.470.2617A} from SDSS-III, and the extended Baryon Oscillation Spectroscopic Survey (eBOSS) \cite{2021PhRvD.103h3533A} from SDSS-IV, have mapped millions of galaxies across vast cosmic volumes, enabling high-precision BAO measurements. Upcoming and ongoing Stage IV spectroscopy surveys, including the Dark Energy Spectroscopic Instrument (DESI) \cite{2016arXiv161100036D}, Euclid \cite{2011arXiv1110.3193L,Euclid_mainpaper}, 4-metre Multi-Object Spectroscopic Telescope (4MOST) \cite{4MOST}, Prime Focus Spectrograph \cite{2014PASJ...66R...1T}, and Wide Field Infrared Survey Telescope \cite{2015arXiv150303757S}, aims to provide multiple sub-percent BAO distance measurements by observing galaxies over a broader redshift range and wider sky coverage. Recently, DESI released BAO measurements\cite{DESI_2024_II,DESI_2024_III,DESI_2024_IV,DESI_2024_V} from its first year of observations, encompassing more than 5.7 million unique galaxy and quasar redshifts within the range $0.1<z<2.1$ as well as Lyman-$\alpha$ forest observations up to a redshift of 3.5. The cosmological interpretation of this data favors a $w_{0}w_{a}$-CDM cosmology, while still being compatible with $\Lambda$-CDM cosmology, if a constant equation of state for dark energy is assumed \cite{DESI_2024_VI,DESI_2024_VII}. This new cosmological tension, also seen in the recent Dark Energy Survey (DES) supernovae results \cite{DES_cosmology_results}, highlights the importance of large-scale BAO surveys to further test these results and refine our understanding of dark energy, while other approaches, such as those proposed in \cite{2023SCPMA..6670413W} and \cite{2023SCPMA..6720412J}, offer additional avenues for precision cosmology.

As another Stage IV galaxy survey, the China Space Station Telescope (CSST) \cite{Zhan2011, Zhan2018, Zhan2021, 2019ApJ...883..203G} is a space-based observatory that will share the same orbit as the Chinese Manned Space Station and is scheduled for launch around 2027. CSST features a 2-meter telescope with a large field of view ($\gtrsim 1.1 \mathrm{deg}^2$) and will cover a total sky area of 17 500 deg$^2$ over a 10-year survey. It has two primary objectives: conducting a photometric imaging survey of billions of galaxies to investigate weak gravitational lensing, and measuring the redshifts of tens of millions of galaxies using slitless spectroscopy to study galaxy clustering and BAO. In the context of CSST studies, \cite{2019ApJ...883..203G} predicted the constraints on cosmological parameters from the CSST weak lensing (WL) and galaxy clustering statistics, demonstrating a significant improvement through the joint analysis of WL, galaxy clustering, and galaxy-galaxy lensing observables. \cite{Miao2023} estimated the constraints on cosmological and systematic parameters from individual probes as well as multiprobe analyses of CSST surveys. \cite{Lin2022} provided forecasts on the sum of neutrino masses based on photo-z galaxy clustering and cosmic shear signals. \cite{2024MNRAS.527.3728D} presented forecasts of the BAO scale measurement from the CSST spec-z and photo-z galaxy clustering, along with their combined analysis. \cite{2024SCPMA..6730411S} investigated the use of CSST galaxy surveys combined with gravitational-wave data to achieve precise measurements of the Hubble constant. \cite{2024SCPMA..6709511J} explored the potential of the CSST ultra-deep field survey to constrain cosmological parameters using superluminous supernovae as standardizable candles. \cite{2023SCPMA..6629511L} provides forecasts of  cosmological constraints from the CSST-like supernova observations (e.g., \cite{2024SCPMA..6719512L}).

The slitless spectroscopy capability of CSST is particularly advantageous for performing large-scale redshift surveys in a highly efficient manner. By capturing light from a wide field without the need for individual slits, CSST will be able to observe vast numbers of galaxies simultaneously, making it well-suited for studying large-scale cosmic structures. However, slitless spectroscopy also presents specific challenges, particularly regarding redshift uncertainties. Unlike slit-based spectroscopy, slitless methods derive redshifts from dispersed light across wide fields, leading to larger redshift errors. These uncertainties directly affect the precision of distance measurements from the BAO signal, especially along the radial direction, where redshift is crucial for determining the line-of-sight distance. Accurate forecasting of the impact of redshift uncertainties on the BAO signal within the context of the CSST is therefore essential to maximize the scientific return of this mission.

One of the most effective tools for detecting the BAO is the two-point correlation function (2PCF), which reveals a characteristic BAO bump at approximately $s \sim 100 \mpch$. While this method has been successful in previous surveys (e.g., \cite{2005ApJ...633..560E, 2005MNRAS.362..505C, 2011MNRAS.416.3017B, 2011ApJ...728...34T, 2017MNRAS.464.1640S, 2017MNRAS.469.1369S, 2018MNRAS.481.3160W}), it is particularly sensitive to redshift uncertainties introduced by slitless spectroscopy. These uncertainties can blur the sharpness of the BAO peak, thereby reducing the accuracy of distance measurements (e.g., \cite{2017MNRAS.472.4456R}). To overcome this limitation, we propose using the wedge 2PCF, which separates galaxy pairs into distinct angular wedges relative to the line of sight. This approach exploits the fact that redshift uncertainties predominantly affect the radial direction, facilitating a cleaner detection of the BAO signal in the transverse direction. While this method has been applied to photometric samples with typical redshift uncertainties of $\sigma_z \geqslant 0.01(1+z)$ (e.g. \cite{2017MNRAS.472.4456R, 2020ApJ...904...69S, 2022MNRAS.511.3965C}), no analysis has yet been conducted for CSST-level uncertainties, which are approximately $\sigma_z \approx 0.005(1+z)$. Given the lower redshift uncertainties in CSST, it is crucial to evaluate how these more precise measurements impact the wedge 2PCF's ability to detect the BAO signal. 

\begin{figure*}
    \centering
	\includegraphics[width=1.5\columnwidth]{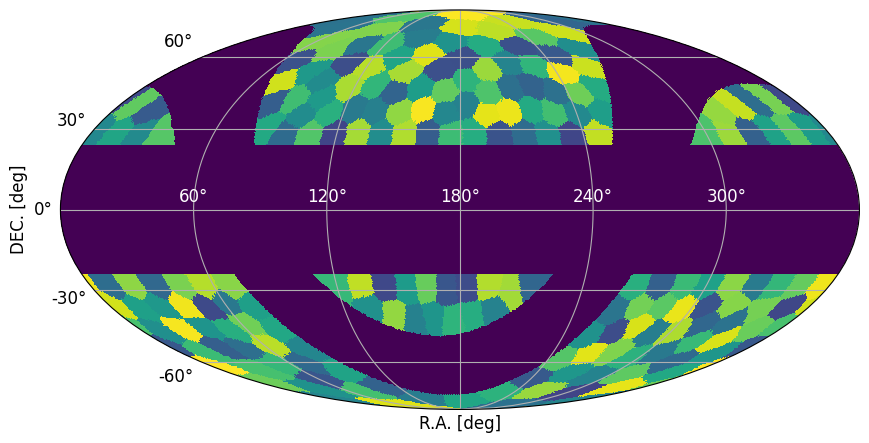}
    \caption{Sky coverage of CSST planned optical survey with color coding to show the 200 subregions for jackknife resampling.}
    \label{fig:ftprint}
\end{figure*}
In this paper, we forecast constraints on angular and radial distances from BAO measurements using a CSST-like redshift survey. Specifically, we aim to assess the effectiveness of the wedge 2PCF in the presence of CSST-level redshift uncertainties and compare its performance to a redshift sample without uncertainties, which serves as an idealized baseline. This comparison allows us to quantify how CSST-level redshift uncertainties impact the precision of BAO measurements. While previous forecasts have typically relied on Fisher matrix techniques (e.g., \cite{2024MNRAS.527.3728D}), which assume ideal conditions, our analysis provides a more realistic estimation by directly incorporating the effects of redshift uncertainties and employing a method better suited to mitigate their impact.

This paper is structured as follows: In Section 2, we introduce the method used to generate the mock CSST galaxy survey. In Section 3, we discuss the theoretical framework for the wedge correlation function and its application to BAO detection. In Section 4, we present our forecast results for the angular and radial distance constraints, and in Section 5, we conclude with a discussion of the implications for cosmology and future survey design.

\section{MOCK CATALOGS}
\label{sec:data}
The simulation of mock galaxy samples for the CSST galaxy survey is a multi-step process involving N-body simulations, halo and subhalo identification, galaxy population algorithms, and survey selection effects. This section outlines the methodology utilized to generate these mock samples, based on the procedures described in the \cite{2024MNRAS.529.4015G}.

The foundation of our mock galaxy samples is the Jiutian N-body simulation, which was run at the High-Performance Computing Center at Kunshan using L-GADGET, a memory-optimized version of the GADGET2 code \cite{2005MNRAS.364.1105S}. The Jiutian simulation describes the distribution of $6144^3$ dark-matter particles in a periodic box of $1000 \mpch$, with a particle mass of $3.723 \times 10^8 h^{-1} M_\odot$. The cosmological parameters were set to be consistent with the Planck2018 results \cite{2020A&A...641A...6P}: $\Omega_m = 0.3111$, $\Omega_{\Lambda} = 0.6889$, $\Omega_b = 0.049$, $\sigma_8 = 0.8102$, and $n_s = 0.9665$.

Dark matter haloes are identified using the friends-of-friends (FOF) algorithm \cite{1985ApJ...292..371D} with a linking length of 0.2 times the mean interparticle separation. Subhaloes and their evolutionary histories are further processed using the Hierarchical Bound-Tracing (HBT+) algorithm \cite{2012MNRAS.427.2437H, 2018MNRAS.474..604H}, which operates in the time domain to track the evolution of each halo throughout the simulation. The minimum number of particles in a subhalo is set to 20, corresponding to a minimum halo mass of approximately $7.5 \times 10^9 h^{-1} M_\odot$. The HBT+ code provides high-quality subhalo catalogues and a robust merger tree by following the most-bound particle when a subhalo is no longer resolved. 

We populate the dark matter halos in our simulation with mock galaxies using the Subhalo Abundance Matching (SHAM) method. This technique establishes a connection between the properties of galaxies and their corresponding subhaloes by assuming a monotonic relationship between these properties \cite{2004ApJ...609...35K, 2004MNRAS.353..189V, 2006ApJ...647..201C, 2012ApJ...752...41Y,2018ARA&A..56..435W}. The positions and velocities of galaxies are assigned by directly inheriting the positions and velocities of their host subhaloes from the N-body simulations. The luminosities are simulated by matching the z-band cumulative galaxy luminosity functions measured from the DESI One-percent survey \cite{2024AJ....168...58D} and year 1 data obtained in \cite{2024ApJ...971..119W} with the cumulative subhalo mass functions derived from the peak mass of all the subhaloes. To reflect the luminosity-subhalo mass relation more realistically, a scatter in the z-band luminosity, $\sigma_\mathrm{log}(L_z)=0.15$ dex (e.g. \cite{2008ApJ...676..248Y}), is added to each galaxy.

Finally, we construct mock galaxy samples following the same footprint as the CSST galaxy survey. We place a virtual observer at the center of the populated simulation box and stack replicas of the box to generate a past light-cone. This process involves selecting snapshots from the N-body simulations that intersect with the light-cone volume at different redshifts, interpolating the positions and velocities of galaxies between snapshots. We then define an $(\alpha, \delta)$ coordinate system and remove all mock galaxies located outside the CSST wide survey region, which is bounded by $|\beta| > 23.43^\circ$ in ecliptic coordinates and $|b| > 15^\circ$ in Galactic coordinates. In this study, we only keep galaxies with z-band magnitude brighter than $m_{\rm z}<21$. Figure~\ref{fig:ftprint} shows the sky coverage of CSST planned optical survey with color coding to show the 200 subregions for jackknife resampling (as described in the next section). 
\begin{figure}[H]
    \centering
	\includegraphics[width=0.9\columnwidth]{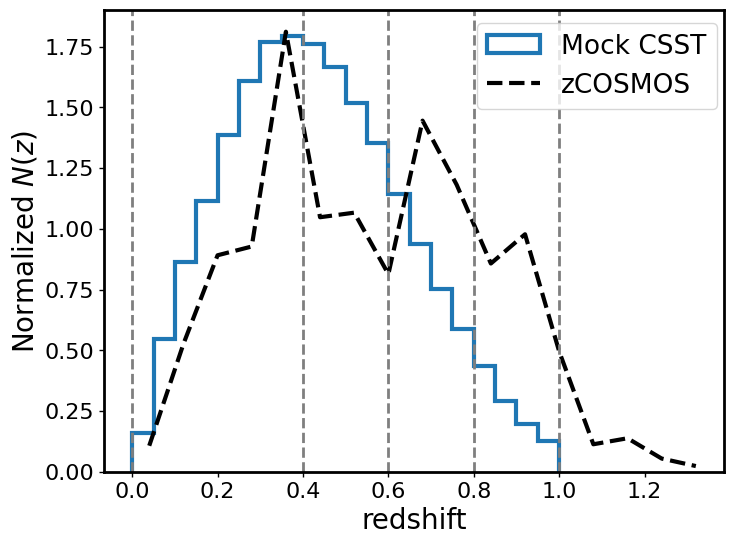}
    \caption{Normalized redshift distribution. Blue solid curve and black dashed line correspond to our mock CSST sample and zCOSMOS catalog, respectively. The vertical
    dashed lines denote the boundaries of the four redshift bins that we divide.}
    \label{fig:nz}
\end{figure}
\begin{figure}[H]
    \centering
	\includegraphics[trim=0cm 6cm 0cm 6cm,clip=True, width=0.9\columnwidth]{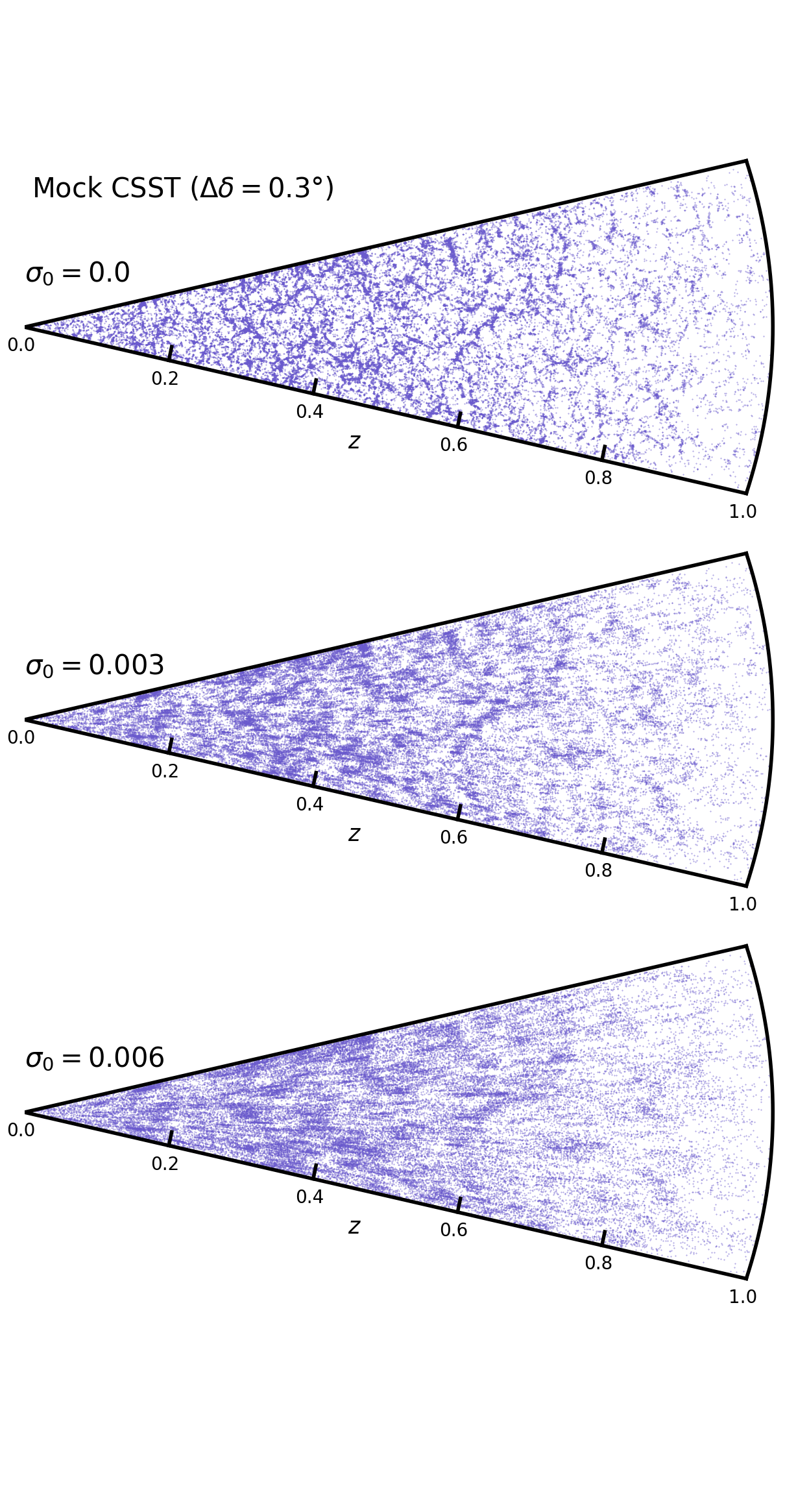}
    \caption{Distribution of the mock CSST galaxy samples  in a 0.3$^\circ$ slice along R.A.. From top to bottom, the panels represent the cases of $\sigma_0=0.0$, $\sigma_0=0.003$, and $\sigma_0=0.006$, respectively. }
    \label{fig:distri}
\end{figure}

Figure~\ref{fig:nz} shows the normalized redshift distributions for our mock CSST sample (blue solid curve). For comparison, we also include the redshift distribution derived from the zCOSMOS catalogue \cite{2007ApJS..172...70L,2009ApJS..184..218L} (black dashed line), which was adopted by \cite{2019ApJ...883..203G, 2024MNRAS.527.3728D} using 20690 galaxies in a 1.7 deg$^2$ field. The redshift distribution of our mock sample generally follows the observed trend. Since the zCOSMOS catalog covers a small fraction of the sky, a significant difference between the two distributions is expected due to cosmic variance. We then incorporate redshift uncertainties expected from the slitless spectroscopy of the CSST, modeled using a Gaussian distribution as,
\begin{equation}
    \sigma_z = \sigma_0(1+z). 
\end{equation}
We consider three scenarios: $\sigma_0 = 0.0$ as a reference case, $\sigma_0 = 0.003$ as a moderate case, and $\sigma_0 = 0.006$ as a pessimistic case. As an illustration, Figure~\ref{fig:distri} compares the mock CSST galaxy sample distributions across the three $\sigma_0$ cases, highlighting the impact of increasing redshift uncertainties on the observed galaxy distribution.  The galaxy sample is further divided into four redshift bins: $0.0 < z \leqslant 0.4$, $0.4 < z \leqslant 0.6$, $0.6 < z \leqslant 0.8$, and $0.8 < z \leqslant 1.0$. The wider first bin is designed to reduce the cosmic variance by increasing the survey volume. Table~\ref{tab:gax_sub} provides the effective redshift, total galaxy count, and number density for each redshift bin.  To generate random samples, we randomize the positions of galaxies before stacking the simulation box to construct the light cone. Geometric and magnitude selection are subsequently applied. To ensure consistent redshift distributions across all $\sigma_0$ cases, the same redshift uncertainty is applied to the random samples.
\begin{table}[H]
\centering
\begin{threeparttable}\caption{Galaxy subsamples}\label{tab:gax_sub}
\begin{tabular}{ccccc}
\toprule
$z$ & $z_\mathrm{eff}$ & $N_\mathrm{gax}$ & $\bar{n}_g10^4$ \\
    &                  &                  & $[h^{3}\mathrm{Mpc}^{-3}]$ \\

\hline
{$0.0 < z \leqslant 0.4$} & ${0.255}$ & ${51\,\,028\,\,414}$ & ${226.279}$ \\
{$0.4 < z \leqslant 0.6$} & ${0.494}$ & ${34\,\,829\,\,517}$ & ${82.843}$ \\
{$0.6 < z \leqslant 0.8$} & ${0.685}$ & ${18\,\,914\,\,076}$ & ${28.988}$ \\
{$0.8 < z \leqslant 1.0$} & ${0.874}$ & ${\,\,\,5\,\,795\,\,176}$ & ${6.763}$ \\\bottomrule
\end{tabular}
\end{threeparttable}
\end{table}
By following these methods, we aim to generate high-fidelity mock samples that can be used to evaluate the performance of the CSST galaxy survey and to understand the underlying cosmology.

\section{METHODOLOGY}

\subsection{The wedge correlation function measurements}

In our analysis, we use 2PCF to reveal the characteristic BAO peak, which corresponds to the preferred separation scale of galaxies due to sound waves in the early universe. 
We estimate the two-dimensional 2PCF, $\xi(s,\mu)$, for galaxies in each sample using the  Landy \& Szalay  estimator \cite{1993ApJ...412...64L}:

\begin{equation}
    \xi(s,\mu) = \frac{{\rm DD(s,\mu)}-2{\rm DR(s,\mu)}-{\rm RR(s,\mu)}}{{\rm RR(s,\mu)}}.
\end{equation}
where DD, RR, and DR are,  respectively, the numbers of
galaxy-galaxy, random-random, and galaxy-random pairs within a spherical shell of radius $s$ and $s + \mathrm{d}s$ and the angular wedge to the LOS enclosed by $\mu$ and $\mu + \mathrm{d}\mu$. The variables $\mu$ is the cosine of the angle between the separation vector ${\bm s}$ and the line-of-sight direction.

The one-dimensional, monopole 2PCF, $\xi(s)$, is estimated by averaging $\xi(s,\mu)$ along costant $s$ using,
\begin{equation}
    \xi(s) = \frac{1}{2}\int^1_{-1}{\xi(s,\mu)d\mu}.
\end{equation}
The monopole 2PCF can reveal the BAO features clearly for the spectroscopic sample. However, the features could be smeared in CSST samples due to redshift measurement uncertainties. To overcome this, we explore the wedge correlation function, $\xi_\mathrm{w}(s,\mu)$, to mitigate the smearing effects in the BAO signal, especially along the line of sight (LOS). The wedge correlation function is given by,
\begin{equation}\label{eq:xis_wedge}
    \xi_{\bar{\mu}}(s) = \frac{1}{\Delta \mu}\int_{\bar{\mu}-\Delta \mu/2}^{\bar{\mu}+\Delta \mu/2}\xi(s,\mu)d\mu.
\end{equation}
It is computed over the interval $\Delta \mu$ centered around $\bar{\mu}$. For our test case here, we then divide $\mu$ into five equal bins with central values of $\bar{\mu} = 0.1$, $0.3$, $0.5$, $0.7$, and $0.9$, and an interval width of $\Delta \mu = 0.2$.

\subsection{The BAO peak fits}
To fit the BAO peak in the wedge 2PCF, we use an empirical model proposed by \cite{2011MNRAS.411..277S}, which accurately predicts the BAO peak position:
\begin{equation}\label{eq:xis_fit}
    \xi(s) = B + \left(\frac{s}{s_0}\right)^{-\gamma}+\frac{N}{\sqrt{2\pi\sigma^2}}\mathrm{exp}\left(-\frac{(s-s_m)^2}{2\sigma^2}\right),
\end{equation}
where $B$, $s_0$, and $\gamma$ model the correlation function on large scales without BAO feature, while $N$, $\sigma$, and $s_m$ characterize the Gaussian feature representing the BAO peak. The parameter $s_m$ is our estimate of the BAO peak position. 

We employ the Markov Chain Monte Carlo (MCMC) technique to estimate the fitting parameters. The constraints on the BAO peak, $s_m$, for the wedge 2PCF are determined after fully marginalizing over all other parameters. We adopt a standard likelihood function, $\mathcal{L} \propto e^{-\chi^2/2}$, where $\chi^2$ is defined as:
\begin{equation}\label{eq:chi2_mcmc}
    \chi^2 = \left[{\bm \xi}^\mathrm{data}(s) - {\bm \xi}^\mathrm{model}(s) \right] 
    \mathbf{C}^{-1}
   \left[{\bm \xi}^\mathrm{data}(s) - {\bm \xi}^\mathrm{model}(s) \right]^{\rm T}.
\end{equation}
\begin{figure}[H]
    \centering
	\includegraphics[width=1.0\columnwidth]{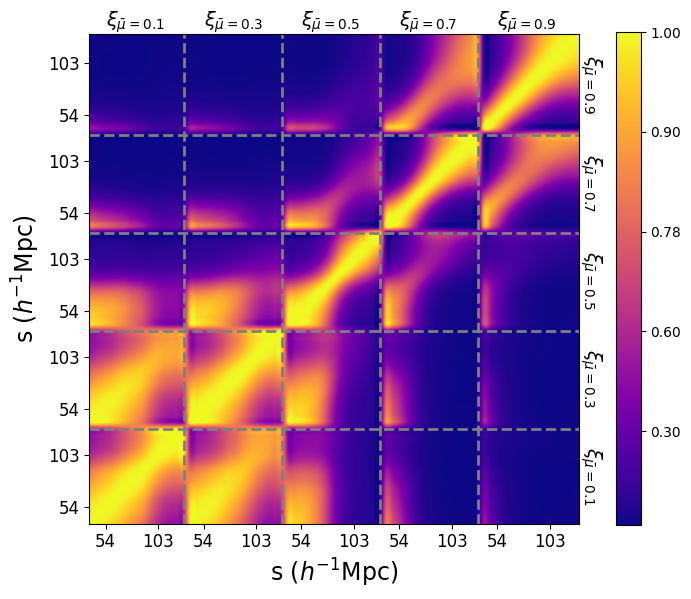}
    \caption{Correlation coefficients of the wedge 2PCF vector $\bmxi^\mathrm{data}$ for galaxy sample within $0.4<z<0.6$ in case of $\sigma_0 = 0.003$. Each grid corresponds to the correlation between two elements of $\bmxi^\mathrm{data}$, as indicated at the top and right margins of the figure.}
    \label{fig:cov_concat}
\end{figure}
\noindent Here, ${\bm \xi}^\mathrm{data} = \left[\xi_{\bar{\mu}=0.1}, \xi_{\bar{\mu}=0.3}, \dots, \xi_{\bar{\mu}_\mathrm{max}} \right]^\mathrm{data}$ represents the vector of measured $\xi_{\bar{\mu}}$ values within the several $\bar{\mu}$ bins, where $\bar{\mu}_\mathrm{max}$ denotes the maximum $\bar{\mu}$ bin utilized. Similarly, ${\bm \xi}^\mathrm{model} = \left[\xi_{\bar{\mu}=0.1}, \xi_{\bar{\mu}=0.3}, \dots, \xi_{\bar{\mu}_\mathrm{max}} \right]^\mathrm{model}$ represents the vector of modeled correlation functions, as defined by Eq.~(\ref{eq:xis_fit}). The matrix $\mathbf{C}^{-1}$ is the inverse covariance matrix between ${\bm \xi}^\mathrm{data}$ and ${\bm \xi}^\mathrm{model}$. In this analysis, we perform the fitting by combining $\xi_{\bar{\mu}}(s)$ obtained from different $\bar{\mu}$ bins, considering that the wedge 2PCFs are expected to exhibit correlations across different $\bar{\mu}$ bins.

In our analysis, we use the jackknife method to estimate the covariance matrix of correlation functions. Figure~\ref{fig:ftprint} shows the footprint divided into $200$ subregions of approximately 87.5 ${\rm deg}^2$ each, based on the method\footnote{\url{https://github.com/rongpu/pixel_partition}} described by \cite{2021MNRAS.501.3309Z}. The covariance matrix is then calculated as,
{\small
\begin{align}
     \mathrm{C}\left[\xi_{\bmu_n}(s_i),\xi_{\bmu_m}(s_j)\right] = \frac{N_{\rm jk}-1}{N_{\rm jk}}\sum_{k=1}^{N_{\rm jk}}\left[\xi_{\bmu_n}^k(s_i)-\bar{\xi}_{\bmu_n}(s_i)\right]\left[\xi_{\bmu_m}^k(s_j)-\bar{\xi}_{\bmu_m}(s_j)\right], 
\end{align}
}
\noindent where $\xi_{\bmu_n}^k(s_i)$ represents the values of the $\bmu_n$-bin wedge 2PCF at the $i$-th separation bins $s$, measured from the $k$-th jackknife sample (similarly for  $\xi_{\bmu_m}^k(s_j)$), $\bar{\xi}$ denotes the mean over all jackknife samples, and $N = 200$ is the number of jackknife resamples. 

However, it has been shown that traditional jackknife estimates of the covariance matrix tend to overestimate the true covariance \cite{2009MNRAS.396...19N, 2016MNRAS.456.2662F}, due to systematic biases in handling cross-pairs between subsamples \cite{2022MNRAS.514.1289M,2024MNRAS.527.9048T}. To address this issue, we applied the correction method proposed by \cite{2022MNRAS.514.1289M}, implemented using {\tt pycorr}\footnote{\url{https://github.com/cosmodesi/pycorr}}. This method introduces a scaling factor to properly weight the contributions of auto- and cross-pairs in jackknife realizations, ensuring that pair counts are consistently scaled across subsamples. Figure~\ref{fig:cov_concat} shows $\mathrm{C}\left[\xi_{\bmu_n}(s_i),\xi_{\bmu_m}(s_j)\right]$ for galaxy sample within $0.4<z<0.6$ for the case of $\sigma_0 = 0.003$. Each grid cell corresponds to the correlation between two elements of $\bmxi^\mathrm{data}$, as indicated at the top and right margins of the figure.

To ensure an unbiased covariance matrix, we adopt the correction proposed by \cite{2022MNRAS.510.3207P} instead of the commonly used Hartlap factor \cite{2007A&A...464..399H}. The corrected covariance matrix is given by,
\begin{align}
    &\mathrm{C}^{\prime}=\frac{(N_{\rm jk}-1)[1+B(N_{\rm bin}-N_{\rm p})]}{N_{\rm jk}-N_{\rm bin}+N_{\rm p}-1}\mathrm{C}, \\
    &B=\frac{(N_{\rm jk}-N_{\rm bin}-2)}{(N_{\rm jk}-N_{\rm bin}-1)(N_{\rm jk}-N_{\rm bin}-4)},
\end{align}
where $N_{\rm bin}$ is the number of data points we are fitting to and $N_{\rm p}$ is the number of free parameters in the model.

\subsection{Theoretical model for the correlation function}

In our analysis, we focus on fitting the BAO peak position rather than the full correlation function. The primary goal of BAO peak fitting is to extract the key cosmological distance information while minimizing the impact of systematic errors and model complexities. This approach is particularly suitable for the CSST slitless spectroscopic sample, where significant redshift uncertainties introduce challenges for traditional full-shape correlation function fitting. The redshift errors in slitless spectroscopy can blur the clustering signal, degrading the precision of redshift-space distortion (RSD) measurements and leading to potential biases. By isolating the BAO peak, which is primarily driven by linear scales and less sensitive to non-linear effects, we can robustly determine cosmological distances without the need to model the entire correlation function.

To determine the theoretical BAO peak position, we first compute $\xi^{\mathrm{th}}(s, \mu)$ in each redshift bin and integrate it over the relevant $\mu$ ranges to obtain the wedge correlation function $\xi^{\mathrm{th}}_{\bar{\mu}}(s)$, corresponding to the same $\bar{\mu}$ as the observational data. The BAO peak is then extracted by fitting $\xi^{\mathrm{th}}_{\bar{\mu}}(s)$ using the model described in Eq.~(\ref{eq:xis_fit}).

 We compute the theoretical correlation function $\xi^\mathrm{th}(s,\mu)$ from the power spectrum as,
\begin{equation}\label{eq:xis_mod}
    \xi^\mathrm{th}(s,\mu) = \frac{1}{(2\pi)^3}\int d^3k P(k,\mu_k) e^{i\bm{k} \cdot \bm{s}},
\end{equation}
where $P(k, \mu_k)$ is the anisotropic redshift-space power spectrum, with $\mu_k$ denoting the cosine of the angle between the wave vector $\bm{k}$ and the line of sight. In our analysis, $P(k, \mu_k)$ is modeled using the Taruya, Nishimichi, and Saito (TNS) model \cite{2010PhRvD..82f3522T}, which accounts for non-linear effects and RSD. The TNS model is given by,
\begin{align}
    P(k,\mu_k) = \Bigl[ \, P_{\delta\delta}(k)
    +2\,f\,\mu^2\,P_{\delta\theta}(k)
    + f^2\,\mu^4\,P_{\theta\theta}(k) \nonumber \\
    + A(k,\mu_k) + B(k,\mu_k) \,\Bigr]\, D^{\rm FoG}\, ,  
\end{align}
where $P_{\delta\delta}(k)$, $P_{\delta\theta}(k)$, and $P_{\theta\theta}(k)$ represent the density-density, density-velocity, and velocity-velocity power spectra, respectively. The terms $A(k, \mu_k)$ and $B(k, \mu_k)$ are the non-linear corrections and are expanded as power series of $\mu_k$. Those spectra are computed using the fiducial cosmological parameters. The Fingers-of-God (FoG) effect is modeled by,
\begin{equation}
    D^{\rm FoG} = \mathrm{exp}\left[-\left(k\,\mu_k\,\sigma_{\rm p}\right)^2\right],
\end{equation}
with $\sigma_{\rm p}$, representing the one-dimensional velocity dispersion, is adopted to be $4.2 \mpch$ in our analysis (e.g. \cite{2019MNRAS.488..295S}). Such value corresponds to approximately 400 km/s, which is consistent with typical values observed in galaxy clustering measurements (e.g. \cite{1998ApJ...494....1J}). Thus, the theoretical correlation function, $\xi^{\text{th}}(s, \mu)$, is parameterized by cosmological distance measures, such as $D_A$ and $H^{-1}$, along with growth functions and velocity dispersion terms. Although these components are essential for modeling clustering, the BAO feature shows only a weak dependence on the growth functions and $\sigma_{\text{p}}$. To address any residual effects, we adopt the TNS model, which incorporates these terms, and account for their influence in the covariance analysis.

It is noted that Equation~(\ref{eq:xis_fit}) is used to extract the BAO peak position from the measured wedge correlation functions, while Equation~(\ref{eq:xis_mod}), incorporating the TNS model, is essential for generating predictions of the theoretical BAO peak. The observed and theoretical peaks are subsequently used to constrain the BAO distance parameters by computing the chi-square statistic, as described in the next subsection.

\subsection{The cosmic distance constraint }\label{sec:mea_dist}

The position of the BAO peak, $s_m$, observed in the wedge correlation function, is directly linked to the comoving sound horizon at the drag epoch, $r_d$, and is influenced by both the angular diameter distance, $D_A(z)$, and the Hubble parameter, $H(z)$.  To constrain $D_A(z)$ and $H(z)$ using $s_m$, we account for the Alcock-Paczynski (AP) effect \cite{1979Natur.281..358A}, which arises due to the assumed fiducial cosmology used to convert galaxy redshifts into distances. The AP effect essentially corresponds to the quadropole moment of the 2PCF. By measuring the the 2PCF in wedges, we are able to capture this additional information beyond the monopole. A straightforward decomposition of the wedge 2PCF enables separate measurements of  $D_A(z)$ and $H(z)$, rather than just their combined effect. If the assumed cosmology deviates from the true cosmology, geometric distortions occur both parallel and perpendicular to the LOS, leading to an anisotropy in the galaxy clustering. To account for these distortions, the cosmology is varied  by the scaling factors,  expressed as,
\begin{equation}
    \alpha_\perp = \frac{D_A(z)\,r_d^{\rm fid}}{D_A^{\rm fid}(z)\,r_d}, \quad \text{and} \quad \alpha_\parallel = \frac{H^{\rm fid}(z)\,r_d^{\rm fid}}{H(z)\,r_d},
\end{equation}
where $D_A^{\rm fid}(z)$, $H^{\rm fid}(z)$, and $r_d^{\rm fid}$ represent the angular diameter distance, Hubble parameter, and comoving sound horizon at the drag epoch in the fiducial cosmology, respectively.

We then construct a grid of $(\alpha_\perp, \alpha_\parallel)$ values covering the range $[0.6, 1.4]$. For each grid point, we calculate the theoretical wedge correlation function $\xi^{\rm th}_{\mu}(s)$ and determine the corresponding BAO peak location $s^{\rm th}_m$. By comparing $s^{\rm th}_m$ with the observed $s^{\rm obs}_m$, we compute the chi-square statistic as:
\begin{equation}
    \chi^2(\alpha_\perp, \alpha_\parallel) = \frac{\left[ s^{\rm th}_m(\alpha_\perp, \alpha_\parallel) - s^{\rm obs}_m \right]^2}{\sigma^2_{s_m}},
\end{equation}
where $\sigma^2_{s_m}$ is the uncertainty in $s^{\rm obs}_m$ obtained from MCMC fitting.

With the constrained scaling factors, the distance observables are given by:
\begin{equation}
    \frac{D_A(z)}{r_d} = \alpha_\perp \frac{D^{\rm fid}_A(z)}{r^{\rm fid}_d},
\end{equation}
and
\begin{equation}
    H(z)r_d = \frac{H^{\rm fid}(z) r^{\rm fid}_d}{\alpha_\parallel}.
\end{equation}

In total, by isolating the BAO peak and modeling the anisotropic correlation function with the AP effect, we can extract cosmological distance measures $D_A(z)$ and $H(z)$. We expect the combination of the scaling factors $\alpha_\perp$ and $\alpha_\parallel$ allows for a robust determination of these observables, even in the presence of redshift uncertainties and geometric distortions.

\begin{figure*}
    \centering
	\includegraphics[width=2.0\columnwidth]{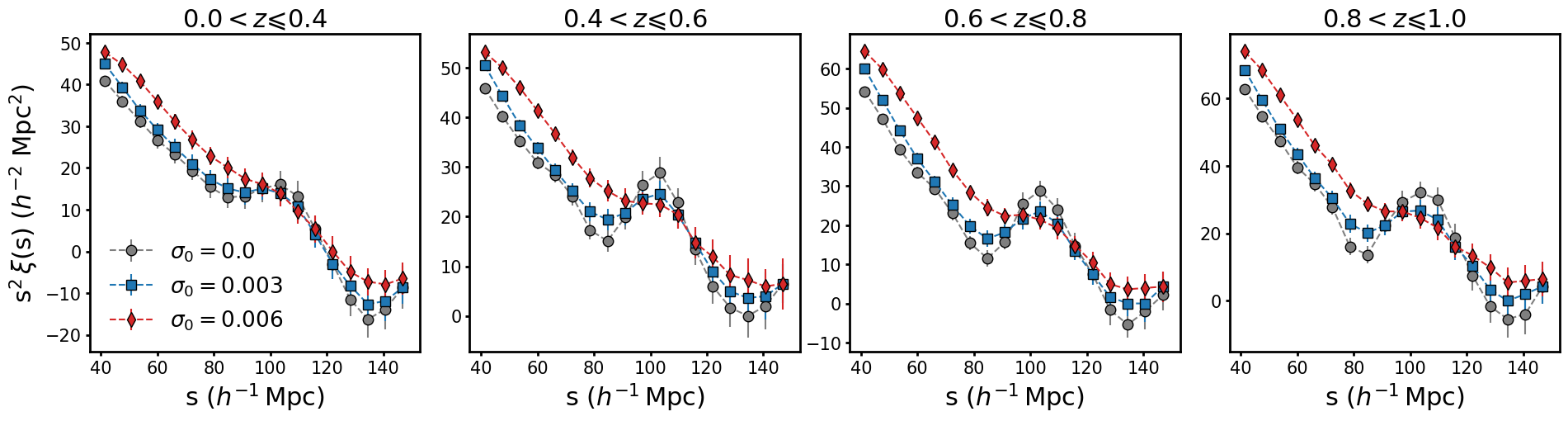}
    \caption{Comparisons of the monopole correlation function for different redshift uncertainty cases. Black circles, blue squares and red diamonds correspond to $\sigma_0=0.0$, 0.003, and 0.006, respectively. Different panels are for different redshift bins, as indicated at the top of each panel. }
    \label{fig:xis_mono}
\end{figure*}
\begin{figure*}
    \centering
	\includegraphics[width=2.0\columnwidth]{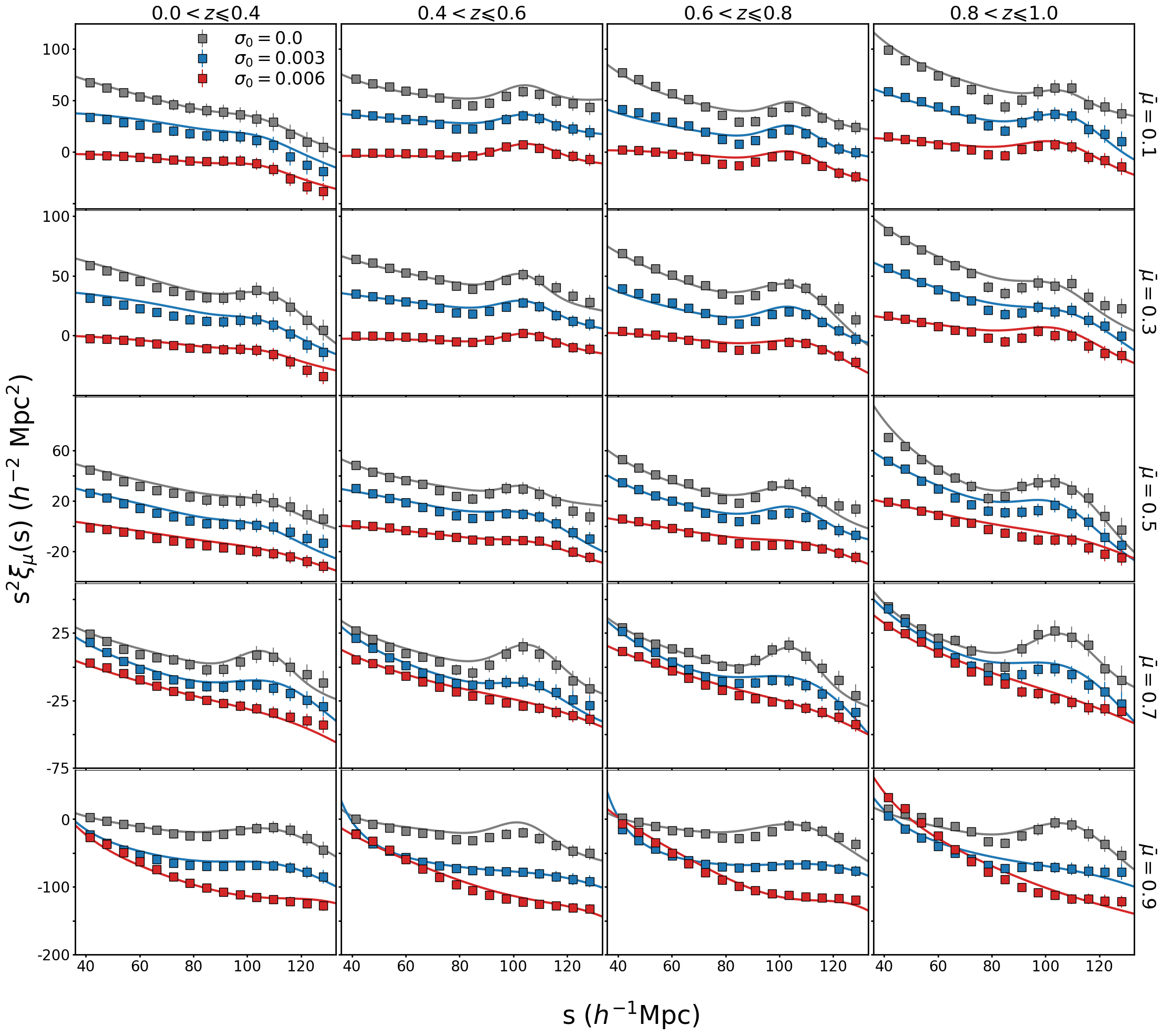}
    \caption{Comparisons of the wedge correlation function for different redshift uncertainty cases. Black circles, blue squares and red diamonds correspond to $\sigma_0=0.0$, 0.003, and 0.006, respectively. Different panels are for different redshift bins, as indicated. The solid color lines in each panel show the best-fitting model obtained from Eq.\ref{eq:xis_fit} by applying the MCMC technique. For clarity, the amplitudes for $\sigma_0=0.003$ and $\sigma_0=0.006$ are shifted downward by subtracting a constant value.}
    \label{fig:xisu}
\end{figure*}
\begin{figure*}
    \centering
	\includegraphics[width=1.8\columnwidth]{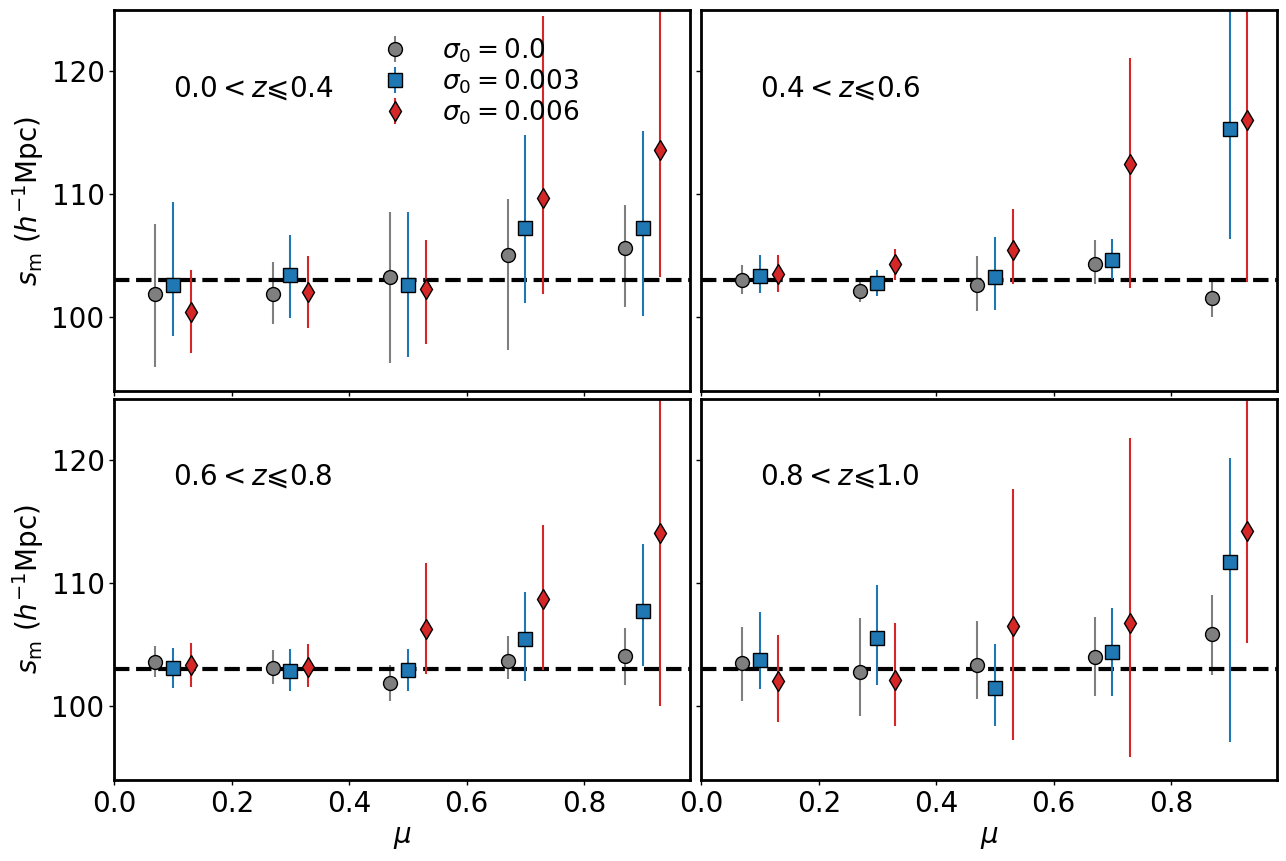}
    \caption{The fitted BAO peak position parameter $s_m$ as a function of $\mu$. Black circles, blue squares and red diamonds correspond to $\sigma_0=0.0$, 0.003, and 0.006, respectively. Different panels are for different redshift bins, as indicated. The back dashed lines are the theoretical values obtained from the {\tt CAMB} based on the fiducial cosmology.}
    \label{fig:sm_fit}
\end{figure*}

\section{Results}
In this section, we present forecasts for the precision of BAO distance measurements using mock CSST galaxy samples. We begin by demonstrating the advantages of the wedge 2PCF over the monopole 2PCF. Next, we fit the BAO peak location, \( s_m \), and use this to constrain the transverse and radial distance scaling factors along with their uncertainties. Our goal is to evaluate the effectiveness of the wedge 2PCF in the presence of redshift uncertainties expected for CSST and to compare its performance against an idealized sample without redshift errors. This comparison enables us to quantify the impact of CSST-level redshift uncertainties on the precision of BAO measurements.

\subsection{Two-point correlation function: monopole vs. wedge}
Figure~\ref{fig:xis_mono} shows the monopole 2PCF for the four different redshift bins, as indicated. Black circles, blue squares and red diamonds correspond to $\sigma_0 = 0.0$, 0.003, and 0.006, respectively. The BAO bump, corresponding to a distinct peak in the correlation function, is typically observed around $s \sim 100 \mpch$. In the absence of redshift uncertainty ($\sigma_0 = 0.0$), the BAO bump is most prominent across all redshift bins. With moderate redshift uncertainty ($\sigma_0 = 0.003$), the BAO bump becomes less distinct as the peak's amplitude decreases and the bump broadens. This blurring effect is caused by redshift errors introducing uncertainties in the comoving distances of galaxies, leading to a smearing of the correlation function and a reduction in the contrast of the BAO signal. Under higher redshift uncertainty ($\sigma_0 = 0.006$), the peak is significantly flattened and broadened, making the BAO feature more difficult to detect.

To address the challenges posed by redshift uncertainties, the wedge 2PCF method proves more effective, as shown in Figure~\ref{fig:xisu}. The different columns correspond to different redshift bins, as indicated at the top of each column, while the rows correspond to different $\mu$ bins, as indicated on the right of each row. The visibility of the BAO bump is expected to vary depending on the redshift uncertainty $\sigma_0$ and the $\mu$ bin. For $\sigma_0 = 0.0$, the BAO bump appears sharp and well-defined across all $\mu$ bins. At $\sigma_0 = 0.003$, the bump remains visible but is slightly broadened and less pronounced, with the best preservation in $\bar{\mu} \leqslant 0.7$ bins. In the case of $\sigma_0 = 0.006$, the bump is still observable in the lower $\mu$ bins but becomes increasingly difficult to discern in the higher $\mu$ bins, particularly the last two.

The monopole 2PCF $\xi(s)$ is an effective tool for detecting the BAO peak, but its effectiveness diminishes as redshift uncertainties increase. In upcoming CSST observations, significant redshift errors may blur the spatial distribution of galaxies, thereby reducing the accuracy of the 2PCF. Consequently, the BAO peak becomes less pronounced, complicating the accurate determination of the BAO scale. The wedge 2PCF divides the galaxy pairs into angular bins relative to the line of sight, effectively separating the radial and transverse components of the correlation function. By focusing on the transverse component, which is less affected by redshift errors, the wedge 2PCF helps maintain the clarity of the BAO signal even in the presence of large redshift uncertainties. Furthermore, the wedge 2PFC allows us to gather additional cosmological information via the quadrupole.
\begin{figure*}
    \centering
	\includegraphics[width=1.1\columnwidth]{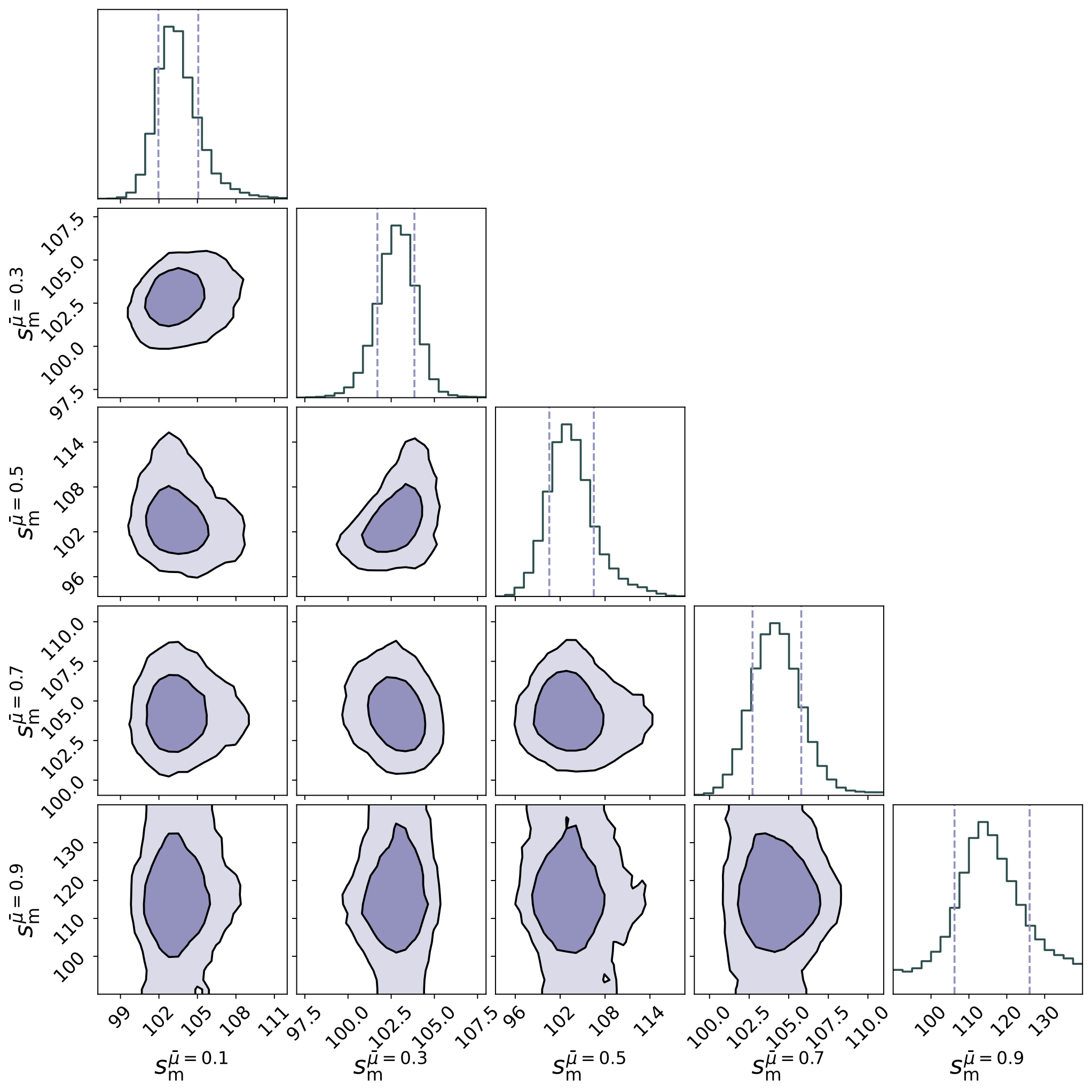}
    \caption{2D confidence contours and 1D marginalized distributions of $s_m$ derived from the MCMC fitting. The results are obtained from the sample within the redshift range $0.4<z\leqslant0.6$ in the case of $\sigma_0=0.003$. The off-diagonal panels display the 2D 68\% and 95\% confidence regions between $s_m$ values for various $\bar{\mu}$ bins (ranging from 0.1 to 0.9), while the diagonal panels illustrate the 1D marginalized distributions for each $\mu$ bin individually. }
    \label{fig:corner}
\end{figure*}
\subsection{Measuring the BAO peak positions}

We applied MCMC fitting to the wedge 2PCFs using the model in Eq.~(\ref{eq:xis_fit}), where we set $\bmu_{\rm max}=0.9$ in Eq.~(\ref{eq:chi2_mcmc}) for all $\sigma_0$ cases. Figure \ref{fig:xisu} shows the best-fitting models as solid colored lines in each panel, demonstrating a good match with the measured data. A prominent Gaussian feature is observed around $s \approx 100 \mpch$ in most $\mu$ bins. However, as $\sigma_0$ increases, this feature becomes less distinct, leading to a flattening of the curves, particularly in higher $\mu$ bins, as seen with the red solid lines in the bottom row panels.  

To further investigate the contamination of pairs along the radial direction, Figure~\ref{fig:sm_fit} shows the fitted BAO peak position parameter, $s_m$, as a function of $\mu$. Black circles, blue squares, and red diamonds correspond to $\sigma_0 = 0.0$, 0.003, and 0.006, respectively. The different panels correspond to different redshift bins, as indicated. As a rough comparison, the black dashed lines represent the theoretical values obtained from {\tt CAMB} based on the fiducial cosmology. For $\sigma_0 = 0.0$, the results are consistent with the theoretical values across all redshifts, with relatively low scatter in the fitted $s_m$ values, indicating good agreement with the expected BAO scale. Introducing $\sigma_0 = 0.003$ slightly shifts the $s_m$ values, leading to some deviation from the theoretical line. The scatter increases compared to the $\sigma_0 = 0.0$ case, particularly at higher $\mu$ values. However, for smaller $\mu$ bins, the fitted $s_m$ values remain relatively consistent with the theoretical prediction, showing minimal scatter. Similarly, in the case of $\sigma_0 = 0.006$, although there are more pronounced deviations from the theoretical prediction at larger $\mu$ bins, with noticeable shifts and larger scatter, the smaller $\mu$ bins exhibit much greater consistency. The fitted $s_m$ values in these lower $\mu$ regions still closely follow the theoretical line, demonstrating minimal deviation even with increased redshift uncertainty. For higher $\mu$ bins, the BAO peaks $s_m$ appear at larger scales.  Such shifts are consistent with the results found in \cite{2017MNRAS.472.4456R, 2022MNRAS.511.3965C,Saulder2024}. As the redshift error increases, the clustering signal is smeared, effectively pushing the peak in the correlation function to larger distances.

Moreover, we observe a minimal correlation of the fitted BAO peak parameter $s_m$ across different $\mu$ bins. As an example, Figure~\ref{fig:corner} presents the MCMC-derived 2D confidence contours and 1D marginalized distributions of $s_m$ for the case of $\sigma_0=0.003$ within the redshift range $0.4<z\leqslant0.6$. The off-diagonal panels display the 2D 68\% and 95\% confidence regions between $s_m$ values for various $\bar{\mu}$ bins (ranging from 0.1 to 0.9), while the diagonal panels illustrate the 1D marginalized distributions for each $\mu$ bin individually. Despite the high correlation observed in the wedge correlation functions, the BAO peak parameter $s_m$ shows minimal correlation across different $\mu$ bins, consistent with the results from \cite{2020ApJ...904...69S}. The weakly correlated 2D confidence contours indicate that the information extracted from the BAO peak location is relatively independent of the specific $\mu$ bin employed. This could be due to the particular $\mu$ binning scheme used, which effectively isolates the information relevant to $s_m$ while minimizing inter-bin correlations. This result highlights the robustness of $s_m$ as a cosmological distance indicator, independent of correlations across different $\mu$ bins, assuming the fiducial cosmology is matched.

Thus, focusing on low-to-intermediate $\mu$ values is expected to provide reliable distance measurements, even in cases of higher redshift uncertainties explored in this study.

\begin{figure*}
    \centering
	\includegraphics[width=2.0\columnwidth]{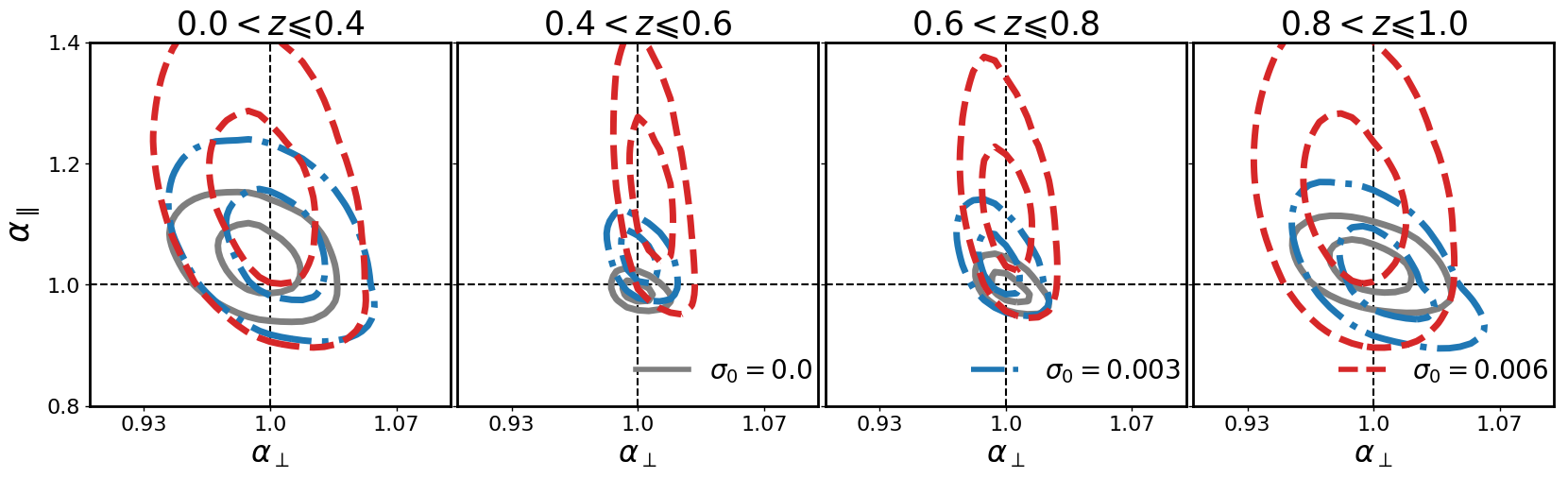}
    \caption{2D confidence contours (with 68\% and 95\% levels) of $\alpha_\perp$ and $\alpha_\parallel$. Different columns correspond to different redshift bins, as indicated at the top of each column.The black solid, blue dash-dotted, and red dashed contours represent the cases for  $\sigma_0 = 0.0$, 0.003, and 0.006, respectively. }
    \label{fig:alpha_con}
\end{figure*}
\begin{figure*}
    \centering
	\includegraphics[width=2.0\columnwidth]{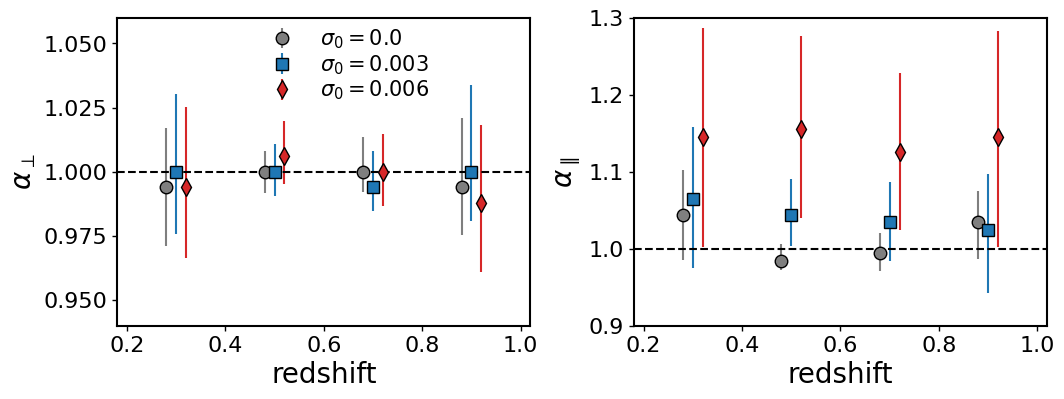}
    \caption{Mean values and 68 per cent CL on $\alpha_\perp$ (left panel) and $\alpha_\parallel$ (right panel) as a function of redshift bins. Black circles, blue squares and red diamonds correspond to the results of $\sigma_0=0.0$, $0.003$, $0.006$, respectively. The black dashed lines represent the fiducial value. }
    \label{fig:alpha_z}
\end{figure*}
\begin{figure*}
    \centering
	\includegraphics[width=2.0\columnwidth]{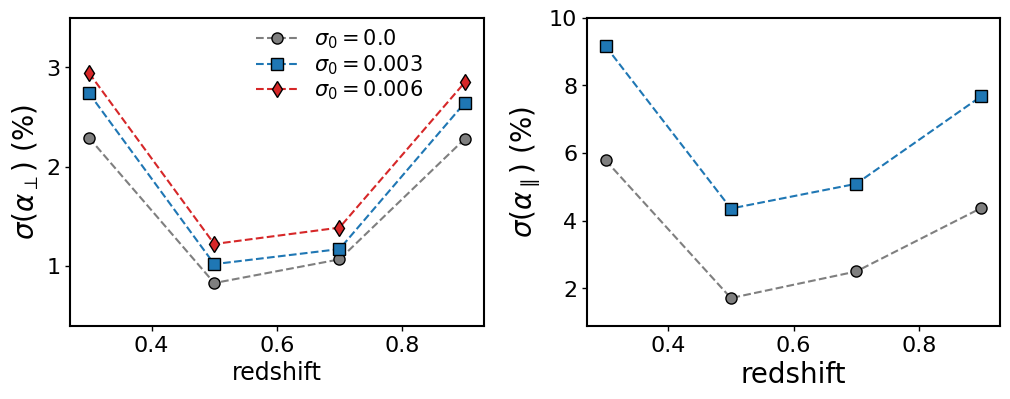}
    \caption{Precision comparison of measured scaling parameters, $\alpha_\perp$ (left panel) and $\alpha_\parallel$ (right panel) across different redshifts. The black circles, blue squares, and red diamonds correspond to results for $\sigma_0 = 0.0$, 0.003, and 0.006, respectively. }
    \label{fig:sigma_alpha}
\end{figure*}
\begin{figure*}
    \centering
	\includegraphics[width=1.5\columnwidth]{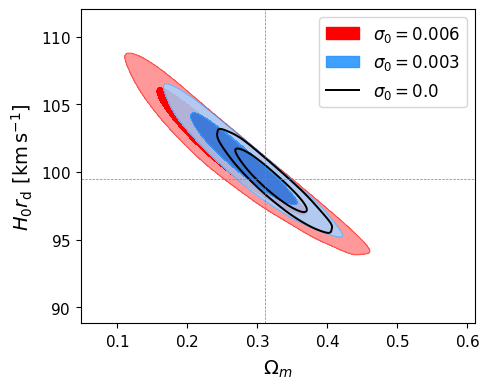}
    \caption{2D confidence contours in the parameter space of $(\Omega_m, H_0 r_\mathrm{d})$. The contours represent the 68\% and 95\% confidence levels. Different colors corresponding to three levels of redshift uncertainty: $\sigma_0 = 0.0$, $0.003$, and $0.006$, as indicated. }
    \label{fig:Om_Hr}
\end{figure*}
\subsection{Measuring the Cosmic Distances}

We use the derived $s_m$ to constrain $\alpha_\perp$ and $\alpha_\parallel$ with the method described in Section~\ref{sec:mea_dist}. Figure~\ref{fig:alpha_con} shows 2D confidence contours at 68\% and 95\% confidence levels for $\alpha_\perp$ and $\alpha_\parallel$. Different columns correspond to different redshift bins, as indicated at the top of each column. The top row shows results obtained by using data from all $\mu$ bins. The black solid, blue dash-dotted, and red dashed contours correspond to cases with $\sigma_0 = 0.0$, 0.003, and 0.006, respectively. Across all redshift bins, $\alpha_\perp$ remains reasonably well-constrained, even when $\sigma_0 = 0.006$. The lack of substantial broadening in the $\alpha_\perp$ direction suggests that $\alpha_\perp$ is relatively insensitive to the value of $\sigma_0$, indicating a robust BAO signal in the direction perpendicular to the line of sight. Although including the biased measurement of $s_m$ in higher-$\mu$ bins of $\sigma_0 = 0.006$, this relative insensitivity is due to $\alpha_\perp$ being dominated by the low-$\mu$ bins, which are less affected by redshift uncertainties. On the other hand, when $\sigma_0 = 0.0$ or 0.003, the contours in the $\alpha_\parallel$ direction remain consistent, showing no significant broadening and indicating a stable constraint on $\alpha_\parallel$ under these conditions. However, with $\sigma_0 = 0.006$, the contours for $\alpha_\parallel$ exhibit reduced precision and biased constraints due to the shifted $s_m$ measurement at higher $\mu$ bins, as shown in Figure~\ref{fig:sm_fit}. This indicates that redshift uncertainties at the level $\sigma_0 = 0.006$ have a more pronounced impact on cosmological parameter estimation from the BAO feature.

Figure~\ref{fig:alpha_z} shows the mean values and 68\% confidence levels for $\alpha_\perp$ (left panel) and $\alpha_\parallel$ (right panel) as a function of redshift bins. The black circles, blue squares, and red diamonds correspond to results for $\sigma_0 = 0.0$, 0.003, and 0.006, respectively. The black dashed lines indicate the fiducial value. For $\alpha_\perp$, the measurements are generally well-constrained around the fiducial value of 1.0 across all redshift bins, regardless of the $\sigma_0$ value. In contrast, the measurements for $\alpha_\parallel$ show a more pronounced impact as $\sigma_0$ increases. For $\sigma_0 = 0.0$ and 0.003, the results remain relatively consistent with the fiducial value, with moderate uncertainties across all redshift bins. However, for $\sigma_0 = 0.006$, the measurements are close to the fiducial value at approximately the $1\sigma$ level but exhibit significant deviations and substantially greater uncertainty.

Figure~\ref{fig:sigma_alpha} shows the precision comparison of measured scaling parameters, $\alpha_\perp$ (left panel) and $\alpha_\parallel$ (right panel) across different redshifts. The black circles, blue squares, and red diamonds correspond to results for $\sigma_0 = 0.0$, 0.003, and 0.006, respectively. The precision of $\alpha_\perp$ remains between 1\% and 3\%, depending on the redshift bin, for both $\sigma_0 = 0.003$ and $\sigma_0 = 0.006$. Compared to $\sigma_0 = 0.0$, the precision of $\alpha_\perp$ decreases by an average of approximately 17\% for $\sigma_0 = 0.003$ and 30\% for $\sigma_0 = 0.006$ across all redshift bins. In contrast, the radial distance constraint $\alpha_\parallel$ shows a more significant reduction in precision, with uncertainties ranging between 4\% and 9\% for $\sigma_0 = 0.003$, depending on the redshift bin.

\subsection{Constraining cosmology}

To further assess the performance of cosmological parameter constraints under the redshift uncertainty expected for the CSST, we derive constraints on cosmology from the measured $\alpha_\perp$ and $\alpha_\parallel$ across all four redshift bins. Figure~\ref{fig:Om_Hr} shows the 2D confidence contours in the parameter space of $(\Omega_m, H_0 r_\mathrm{d})$. The contours represent the 68\% and 95\% confidence levels and are illustrated in different colors corresponding to three levels of redshift uncertainty: $\sigma_0 = 0.0$, $0.003$, and $0.006$, as indicated. An increase in $\sigma_0$ leads to broader contours, clearly showing the degradation of parameter constraints due to increasing redshift uncertainty.

Despite this broadening, the BAO measurements continue to provide meaningful constraints on $\Omega_m$ and $H_0 r_\mathrm{d}$. For $\sigma_0 = 0.003$, the constraints remain relatively robust, suggesting that moderate redshift uncertainties have a limited impact on the precision of BAO-based cosmological measurements. Even at $\sigma_0 = 0.006$, where the contours significantly expand, the results retain their utility in breaking parameter degeneracies and offering insights into the underlying cosmological model. This demonstrates the robustness of BAO as a cosmological probe, even under non-negligible observational uncertainties expected for CSST.

\section{Discussion}

In this study, we employed the empirical fit model from \cite{2011MNRAS.411..277S} to analyze the BAO signal for the CSST-like galaxy survey. While this method is widely used for high-quality spectroscopic data in isotropic cases, its application to data with intermediate redshift uncertainties, such as those expected from the CSST, introduces several challenges. Recent works (e.g., \cite{2022MNRAS.511.3965C}) have highlighted the impact of larger redshift uncertainties (such as for photometric redshifts) on the BAO signal, particularly the ``peak shift'' phenomenon. Specifically, uncertainties in the range we consider here place the analysis in a transition regime where traditional spectroscopic methods struggle to accurately constrain $ \alpha_\parallel $, while photometric approaches relying on binning in $ s_\perp $ instead of $s$ alone are not entirely suitable. These systematic effects could potentially bias the inferred cosmological parameters.

Furthermore, the use of an empirical model introduces limitations in capturing the anisotropic features of the BAO signal, especially at higher $\mu$-bins where the correlation function can dip below zero. As shown in recent analyses (e.g., \cite{2023MNRAS.525.5406M}), template-fitting methods based on simulations or theoretical models provide a more robust framework for dealing with the degradation of the correlation function at these scales. However, the development of accurate templates requires precise modeling of the redshift uncertainty distribution and significant computational resources, which were beyond the scope of this initial study. Notably, the exact redshift accuracy of the CSST remains uncertain, further complicating the development of templates optimized for this survey.

Despite these limitations, our current methodology serves as a proof of concept, demonstrating the potential of the CSST to constrain BAO over a wide redshift range. By focusing on simple empirical models, we provide a baseline for future analyses, acknowledging the challenges and systematic biases that may arise. As the exact redshift accuracy of the CSST is refined and computational resources become available, the adoption of more sophisticated techniques, such as template fitting with detailed modeling of redshift uncertainties, will likely lead to improved constraints on cosmological parameters. This future work will build upon the insights gained here to fully exploit the capabilities of the CSST for precision cosmology.

\section{Summary and Conclusion}

In this work, we conducted a detailed forecasting analysis of BAO distance measurements using the mock CSST galaxy redshift survey. The primary objective was to evaluate the impact of redshift uncertainties on the precision of both angular and radial distance estimates. We employed a method based on the 2PCF using the wedge approach, which separates galaxy pairs into distinct angular wedges. This technique was selected to minimize the effects of larger redshift uncertainties on the extraction of the BAO signal and to maximize cosmological information by including the quadrupole moments of the 2PCF. We applied this approach to mock CSST samples across four redshift bins: $0.0 < z \leqslant 0.4$, $0.4 < z \leqslant 0.6$, $0.6 < z \leqslant 0.8$, and $0.8 < z \leqslant 1.0$. Redshift uncertainties were considered with values $\sigma_0 = 0.003$ and $\sigma_0 = 0.006$ to simulate the expected redshift errors from the slitless spectroscopy in the CSST survey, and the results were compared with an idealized case of no redshift uncertainties.

We find that the BAO peak in the monopole 2PCF $\xi(s)$ is slightly less distinct for $\sigma_0 = 0.003$ and significantly blurred for $\sigma_0 = 0.006$, compared to the ideal case of $\sigma_0 = 0.0$. To address the impact of redshift uncertainties, we demonstrate that the BAO peak remains visible in the wedge 2PCF, with the best preservation occurring at $\bar{\mu} \leqslant 0.7$ for $\sigma_0 = 0.003$, and at $\bar{\mu} \leqslant 0.5$ for $\sigma_0 = 0.006$. By fitting the wedge 2PCFs concatenated from different $\mu$ bins, we confirmed the robustness of our method in estimating the BAO peak positions, $s_m$, across the four redshift bins for all values of $\sigma_0$. The fitted $s_m$ values in the lower $\mu$ bins closely match theoretical predictions, showing minimal deviation even with increased redshift uncertainties. However, more significant deviations from the theoretical predictions are observed at larger $\mu$ bins, with noticeable shifts and increased scatter.

We then use the derived $ s_m $ values to constrain the transverse and radial distance scaling factors, $ (\alpha_\perp, \alpha_\parallel) $. The results indicate that $\alpha_\perp$ measurements are generally well-constrained around the fiducial value of 1.0 across all redshift bins, regardless of the $\sigma_0$ value. The precision of $\alpha_\perp$ remains between 1\% and 3\%, depending on the redshift bin, for both $\sigma_0 = 0.003$ and $\sigma_0 = 0.006$. Compared to the ideal case of $\sigma_0 = 0.0$, the precision of $\alpha_\perp$ decreases by an average of approximately 17\% for $\sigma_0 = 0.003$ and 30\% for $\sigma_0 = 0.006$ across all redshift bins. In contrast, $\alpha_\parallel$ measurements are more sensitive to increases in $\sigma_0$. For $\sigma_0 = 0.003$, the results remain close to the fiducial value, with uncertainties ranging between 4\% and 9\%; for $\sigma_0 = 0.006$, significant deviations from the fiducial value are observed.

Our results demonstrate that redshift uncertainties degrade the precision of distance measurements, particularly in the radial direction. Nevertheless, the wedge method achieves few-percent precision in transverse BAO measurements, proving effective in mitigating the impact of redshift uncertainties. A promising approach to further improving these results is the application of BAO reconstruction techniques, which can recover the original, unperturbed density field, sharpening the BAO signal and enhancing both radial and transverse distance measurements \cite{2007ApJ...664..675E}. In other surveys, reconstruction has been shown to reduce errors successfully (e.g. \cite{2012MNRAS.427.2132P, 2014MNRAS.440.2222T, 2014MNRAS.441.3524K, 2014MNRAS.441.3524K, 2017MNRAS.470.2617A, 2020MNRAS.498.2492G,2024SCPMA..6719513W,2024arXiv240403005P,2024arXiv241119738C}), and applying this method to CSST data could significantly enhance the precision of BAO constraints \cite{2024MNRAS.527.3728D}. Additionally, while the current method focuses solely on the BAO peak position, further improvements may be obtained by employing full-shape fitting techniques (e.g. \cite{2020MNRAS.498.2492G}). Full-shape analyses leverage the entire power spectrum or correlation function, capturing more cosmological information and potentially tightening parameter constraints beyond what is achievable with the BAO peak alone. Both improvements require careful validation in the presence of redshift uncertainties. We will conduct detailed analyses in subsequent papers.

Additionally, we extended the analysis to constrain cosmological parameters using the derived scaling factors $(\alpha_\perp, \alpha_\parallel)$. The results indicate that, although redshift uncertainties reduce the precision of radial distance constraints, the CSST-like galaxy survey still provides robust measurements of $(\Omega_m, H_0 r_\mathrm{d})$, highlighting its potential to advance cosmological studies. Its extensive survey volume and broad sky coverage enable precise measurements of $D_A(z)$, offering critical insights into the universe's expansion history. Although radial distance constraints through $H(z)$ are more sensitive to redshift errors, transverse measurements help to break degeneracies in cosmological models and complement other observational probes. Moreover, the ability of the CSST to independently validate its results against measurements from other galaxy surveys is pivotal for mitigating systematic uncertainties. Variations in galaxy selection, survey design, and redshift determination across different surveys introduce potential systematics that must be rigorously addressed. By providing independent measurements of $D_A(z)$ and $H(z)$, even with moderate redshift uncertainties, the CSST offers a valuable cross-check against other large-scale structure surveys, enabling tighter constraints on cosmological parameters and resolving potential tensions between survey results.

\Acknowledgements{This work is supported by the National SKA Program of China (2022SKA0110200 and 2022SKA0110202),  the National Key R\&D Program of China (2023YFA1607800, 2023YFA1607802, 2023YFA1607804, 2022YFF0503400),  the National Natural Science Foundation of China (Nos.12103037 and 12273020), 111 project No. B20019, and Shanghai Natural Science Foundation, grant No. 19ZR1466800. We acknowledge the science research grants from the China Manned Space Project (Grant No. CMS-CSST-2021-A02, CMS-CSST-2021-A03, CMS-CSST-2021-B01) and the Fundamental Research Funds for the Central Universities (Grant No. XJS221312). X.L. was supported by Science Research Project of Hebei Education Department No. BJK2024134. We acknowledge the Beijing Super Cloud Center (BSCC) for providing HPC resources that have contributed to the research results reported within this paper. This work is also supported by the High-Performance Computing Platform of Xidian University. }

\InterestConflict{The authors declare that they have no conflict of interest.}

\bibliographystyle{unsrt85}
\bibliography{ref}







\end{multicols}
\end{document}